\documentclass[fleqn,usenatbib]{mnras}

\usepackage{newtxtext,newtxmath,multirow,appendix,tabularx,subfigure,booktabs}
\usepackage[T1]{fontenc}
\DeclareRobustCommand{\VAN}[3]{#2}
\let\VANthebibliography\thebibliography
\def\thebibliography{\DeclareRobustCommand{\VAN}[3]{##3}\VANthebibliography}

\usepackage{graphicx}	
\usepackage{amsmath}	

\title[A new class of variability in GRS 1915+105]{A new variability pattern in GRS 1915+105 with NICER and \textit {Insight}-HXMT observations}

\author[Zhihong Shi et al.]{
Zhihong Shi$^{1,2}$,
Qingwen Wu$^{2}$\thanks{E-mail: qwwu@hust.edu.cn},
Zhen Yan$^{3}$,
Bing Lyu$^{4}$
and Hao Liu$^{5}$,
\\
$^{1}$Department of Physics, The University of Hong Kong, Pokfulam Rd., Hong Kong, China\\
$^{2}$School of Physics, Huazhong University of Science and Technology,1037 Luoyu Road, Wuhan, 430074, China\\
$^{3}$Shanghai Astronomical Observatory, Chinese Academy of Sciences, 80 Nandan Road, Shanghai, 200030, China\\
$^{4}$Kavli Institute for Astronomy and Astrophysics, Peking University, Beijing 100871, China\\
$^{5}$University of Science and Technology of China, No.96, JinZhai Road Baohe District, Hefei, Anhui, 230026, China
}

\date{Accepted XXX. Received YYY; in original form ZZZ}

\pubyear{2022}

\begin{document}
\label{firstpage}
\pagerange{\pageref{firstpage}--\pageref{lastpage}}
\maketitle

\begin{abstract}
We explore the timing and spectral properties of GRS 1915+105 based on X-ray observations of NICER and \textit{Insight}-HXMT during the long outburst from 2017 to 2021. We find a new class of variability in the rising stage of the outburst that differs from the formerly reported patterns of light curves. This new variability pattern, which we name class $\psi$, is characterized by several periodic mini pulses superposed on another longer periodic pulse. The periods are $\sim$130 seconds and  $\sim$10 seconds for the main pulses and mini pulses respectively based on the analysis of power spectrum density (PSD) and step-wise filter correlation (SFC), where the SFC method has an advantage in finding the superimposed periodic components. The mini pulses become weak or disappear when the luminosity increases and the light curves change into the classical class $\kappa$. 
The class $\psi$ shows a softer spectrum with lower count rates compared to the class $\kappa$ during the main pulse. The new class $\psi$ shows peculiar timing and spectral properties compared to those of classic class $\kappa$, which can help us to explore the class transition mechanism in GRS 1915+105.

\end{abstract}

\begin{keywords}
accretion, accretion discs – black hole physics – X-rays: binaries – stars: individual: GRS 1915+105.
\end{keywords}

\section{Introduction}


GRS 1915+105 is an extremely variable black hole X-ray binary (BH XRB), exhibiting high amplitude, highly structured variability on a wide range of timescales from milliseconds to a couple of weeks \citep{remillard1997multifrequency}. It was first discovered in 1992 \citep{1994Discovery} and is composed of a fast-spinning BH with a mass of about 12.2 $\rm M_{\odot}$ and a 0.8 $\rm M_{\odot}$ K-M III companion star \citep{reid2014parallax}. The X-ray luminosity of GRS 1915+105 roughly reaches 10–100\% of the Eddington luminosity over the last three decades from 1992 \citep{castro1994discovery}, where the luminosity suddenly decreased by an order of magnitude from mid-2018 \citep{negoro2018maxi,motta2019ami}. \citet{belloni2000model} proposed a classification for the different variability classes of GRS 1915+105, which included 12 different patterns ($\alpha$, $\beta$, $\gamma$, $\delta$, $\theta$, $\kappa$, $\lambda$, $\mu$, $\nu$, $\rho$, $\phi$, and $\chi$), based on the light curves and hardness ratios. Later on, two new variability patterns of $\xi$ and $\omega$ were further reported \citep{naik2002fast,hannikainen2005characterizing}. The peculiar timing and spectral properties of these different patterns make the behavior of GRS 1915+105 particularly difficult to explore in physical detail. Even in the low-luminosity stable class $\chi$, it is still different from the normal hard state as found in many other BH XRBs \citep{massaro2010complex}.  \citet{belloni1997unified} found that the longer duration of the low state will lead to a longer outburst event in the class $\kappa$. Among these different patterns, the most famous one is class $\rho$ (also called heartbeat state), which has very regular and stable pulses. This special pattern is also found in BH XRB of IGR J17091–3624 \citep{capitanio2012peculiar}. \citet{belloni1997unstable} proposed that these different variabilities are possibly caused by the thermal-viscous instabilities with the appearance or disappearance of an inner cold accretion disk. Recently, \citet{massaro2020non} presented a non-linear mathematical model to explain the variability classes and claimed that the change of some parameters can cause the state transition from quiescent class like $\phi$ and $\chi$ to the class $\delta$ and the spiking class $\rho$.

BH XRBs normally show different spectral states at different intensities (e.g., low/hard state and high/soft state). The cold disk may extend to the innermost stable orbit in a high/soft state, while the inner cold disk will transit to hot advection dominated accretion flow in the low/hard state \citep{shakura1973reprint,yuan2014hot}. For GRS 1915+105, three distinct spectral states were defined based on the intensity and hardness ratio: State A, B, and C \citep{belloni2000model,tanaka1996x,remillard2006x}, where two soft states are characterized by strong thermal emission (State A and B with different intensities) and one hard state is dominated by non-thermal emission (State C). GRS 1915+105 was also  the first Galactic source known to exhibit apparent superluminal motion of the relativistic jet \citep{mirabel1994superluminal}. The sub-relativistic steady jet and the wind are also found in different states of GRS 1915+105, which should be closely correlated with the accretion process \citep{mirabel1994superluminal,zoghbi2016disk}. Even though the variability patterns in GRS 1915+105 are different from other BH XRBs, the steady radio emission can be observed in low/hard state, which suggests that despite some differences in the spectral shape, the physics of the jet formation is probably the same for all BH XRBs \citep{fender2004grs}. The intensity variabilities, spectral variations, and different types of outflow allow us to explore the possible evolution of the accretion flow and outflow formation.

In this work, we report a discovery of the new variability pattern based on the NICER and \textit{Insight}-HXMT during the long outburst just before the decay in mid-2018. In Section 2, we present the data reduction of \textit{Insight}-HXMT and NICER, timing and spectral analysis results are shown in Section 3, conclusion and discussion are presented in Section 4.

\section{Data Reduction}

We reduce NICER data for GRS 1915+105 from 28 June 2017 to 27 May 2021 (MJD 57932-59361), which roughly covers the recent outburst (total of 410 observations). The data is reduced using NICER software {\tt\string NICERDAS v06a} distributed in {\tt\string HEAsoft v6.26} with the latest calibration files {\tt\string CALDB v20200722}. We follow the standard NICER calibration processes {\tt\string nicerl2} with standard filter criteria to generate cleaned event files. In the study of hardness ratio, {\tt\string nicerclean} is adopted to generate 0.2--3 keV, 3--6 keV, and 6--12 keV cleaned event data. We use {\tt\string XSELECT V2.4g} to extract the light curves with a time resolution of $1/128$ seconds per bin and the spectra, and then generate power spectrum density (PSD) with {\tt\string powspec}. We adopt {\tt \string nicerarf} and {\tt \string nicerrmf} to obtain the Ancillary Response Files (ARFs) and Response Matrix Files (RMFs) with real-time locations of GRS 1915+105. The {\tt \string space weather} model is applied to generate a background for the spectra. We fit the spectra with {\tt\string Xspec 12.11.1} and PSDs using the method described in \cite{ingram2012modelling}.

The $Insight$-Hard X-ray Modulation Telescope (HXMT) is China's first X-ray astronomical satellite \citep{zhang2020overview}. It was launched on June 15, 2017, and has three collimated telescopes: Low Energy (LE, 1-10 keV), Medium Energy (ME, 5-30 keV), and High Energy (HE, 20-250 keV) X-ray Telescopes \citep{chen2020low,cao2020medium,liu2020high}. \textit{Insight}-HXMT has started to observe GRS 1915+105 since July 2017.  We perform our analysis with observations from 19 July 2017 to 23 June 2019 (MJD 57953-58657), using \textit{Insight}-HXMT Data Analysis Software ({\tt\string HXMTDAS v2.04}) \footnote{http://hxmt.org/software.jhtml}. We select the good time interval (GTI) with recommended criteria: elevation angle (ELV) greater than $10^\circ$,  point offset angle less than $0.1^\circ$, geometric cutoff rigidity (COR) greater than 8 GeV, and data at least 300 seconds away from South Atlantic Anomaly (SAA) area. The backgrounds of the energy spectra and light curves are estimated by using the official independent tools: {\tt\string lebkgmap}, {\tt\string mebkgmap} and {\tt\string hebkgmap} \citep{guo2020background,liao2020background2,liao2020background1}. For timing analysis, {\tt\string helcgen}, {\tt\string melcgen} and {\tt\string lelcgen} are used to generate light curves.

\section{Data Analysis and Results}
\subsection{Timing analysis}
\begin{figure*}
	\includegraphics[width=\textwidth]{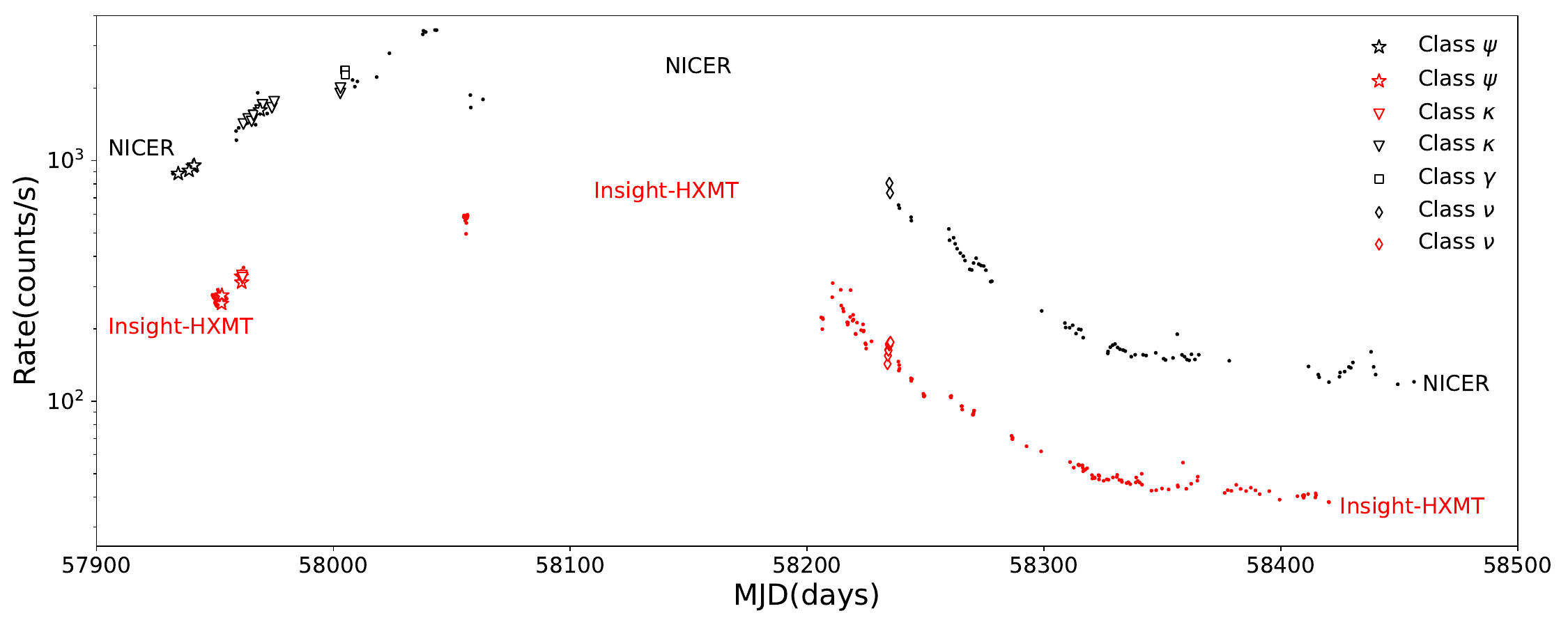}
	\caption{Light curves of GRS 1915+105 as observed by NICER (1--10 keV, black) and \textit{Insight}-HXMT (LE telescope, 1--10 keV, red) from MJD 57900 to 58500, where different classes are marked as different shapes and the new class of variability pattern is marked as the hollow pentagram.}
	\label{fig:lc1}
\end{figure*}

\begin{figure*}
\centering
	\includegraphics[height=0.95\textheight]{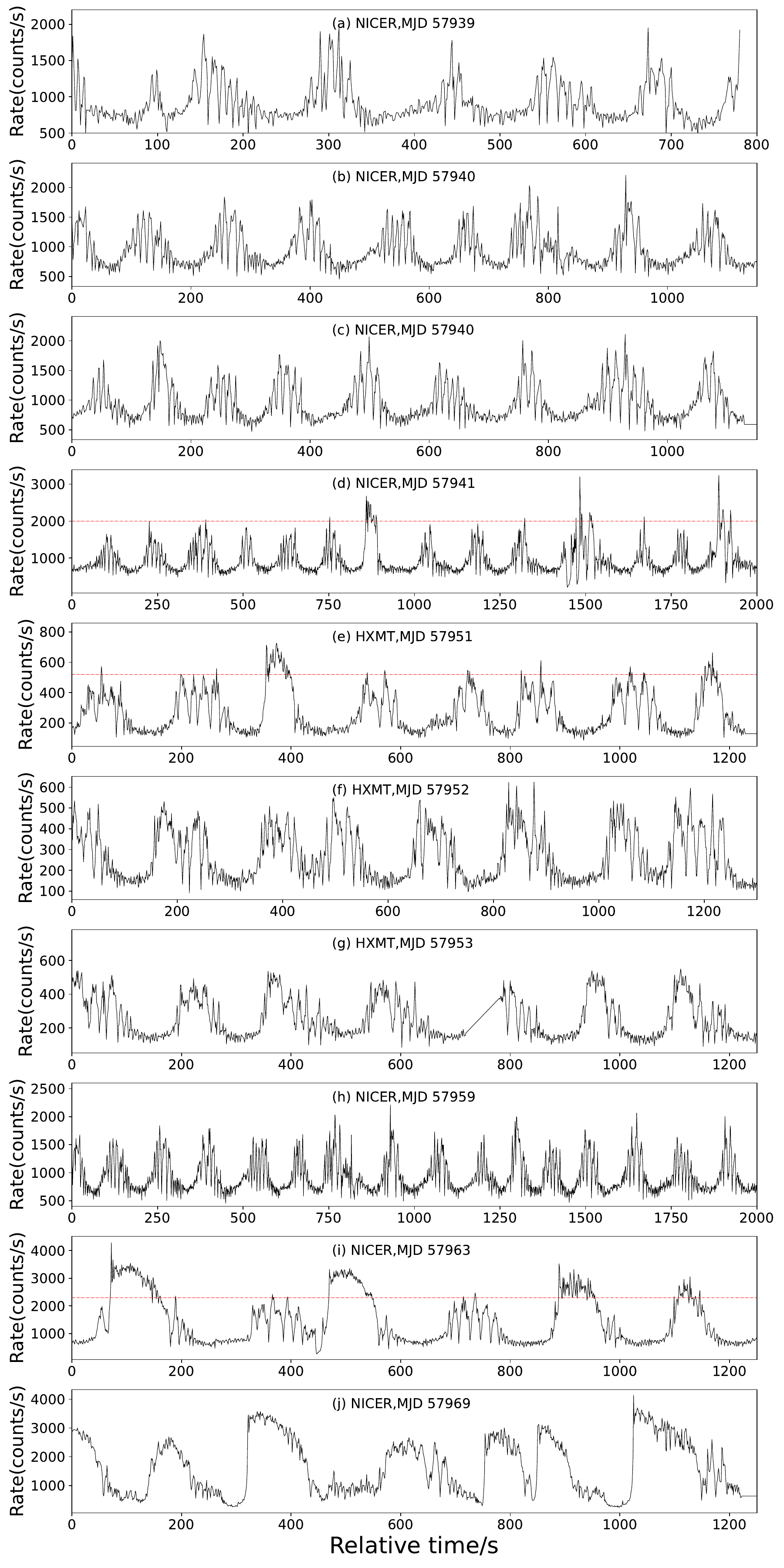}
    \caption{The light curves of the new class $\psi$ (panels a-h) and the class $\kappa$ (panels i and j) as observed by NICER and \textit{Insight}-HXMT with 1s time resolution. The mini pulses are suppressed in panels d, e, and i. }
    \label{fig:lc2}
\end{figure*}

We present the long-term light curves for the selected data from NICER (1--10 keV) and \textit{Insight}-HXMT (1--10 keV of LE detector) in Figure \ref{fig:lc1}, where each data point corresponds to one observation. We classify the variability patterns for the light curves of each observation by visual inspection and find that most observations during this period stay in the stable class $\chi$. We find that 19 NICER observations show different variability patterns, which include 10 observations in class $\kappa$, 2 observations in class $\gamma$, 2 observations in class $\nu$, and 5 observations in a new class that differ from formerly reported patterns (Table \ref{tab:obID}). There are 13 \textit{Insight}-HXMT observations in the new class as found in NICER, 3 observations in class $\kappa$, and 6 observations in class $\nu$. The different classes are shown in Figure \ref{fig:lc1}, where we can find that different classes appear in different outburst stages. We further present the typical light curves (time resolution of 1s per bin) of the new class in Figure \ref{fig:lc2}, which is mainly found in the initial rise stage of the long outburst from MJD 57934 to 57941 (see Figure \ref{fig:lc1}). The new pattern is characterized by several mini pulses ($\sim$10 seconds) superposed on the main pulse ($\sim$130 seconds; see Figure \ref{fig:lc2}). Based on the profile of light curves for this new class, we define it as class $\psi$, because the alphabet $\psi$ seems like several branches grow from a main stem. It should be noted that the light curves from MJD 57941 to MJD 57961 show a mixture of the classes of $\psi$ and $\kappa$, where the mini pulses disappear in some main pulses with count rates higher than a critical value (see the red dot line in Figure \ref{fig:lc2}). With the increasing of X-ray fluxes, the light curves roughly transit to class $\kappa$ in MJD 57961 (bottom panel of Figure \ref{fig:lc2}), where mini pulses are very weak or disappear.

\begin{figure*}
    \centering
	\includegraphics[width=0.8\textwidth]{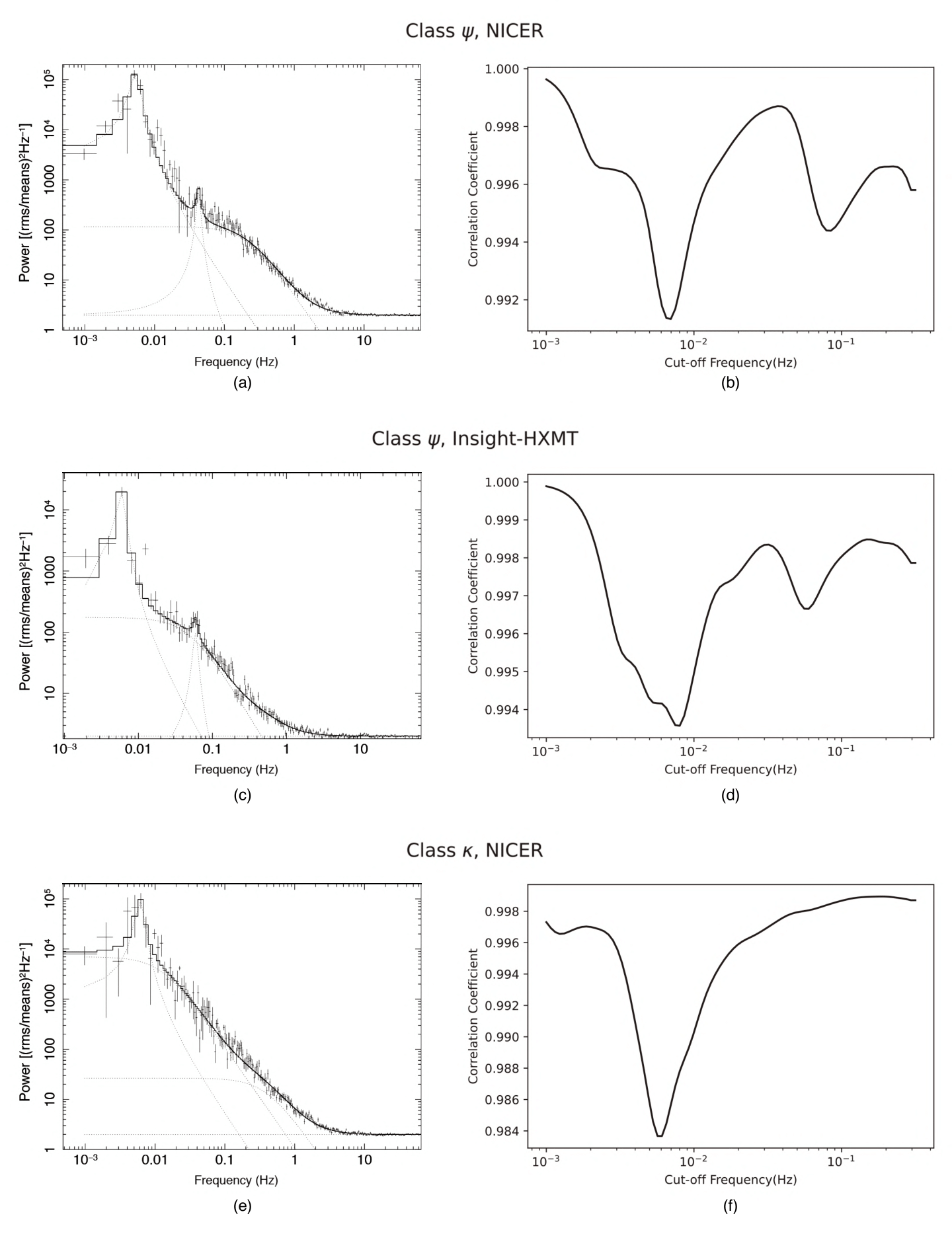}
    \caption{Three examples of PSD (left) and SFC (right) analyses on the class $\psi$ [top panel and middle panel are observed with NICER (ObsID 1103010102) and \textit{Insight}-HXMT(ObsID P010133010105 ) respectively ] and class $\kappa$ (ObsID 1103010112) as observed by NICER. The dotted lines in the left panels are the fittings with Lenrentz functions.}
    \label{fig:psd}
\end{figure*}

To explore the timing properties of this new class and compare them with those of the class $\kappa$, we calculate the PSDs using a time resolution of 1/128 second for the NICER and \textit{Insight}-HXMT observations of class $\psi$ and $\kappa$. In the left panel of Figure \ref{fig:psd}, we present the examples of PSDs for two observations of class $\psi$ and one for class $\kappa$, where the PSDs are fitted with one power-law and 3 or 4 Lorentz functions. Figure \ref{fig:psd} (a) show the PSDs of class $\psi$ light curves observed by NICER (MJD 57959), where the QPO frequencies are 5.29 $\times10^{-3}$ Hz  and 4.25 $\times {10^{-2}}$ Hz (corresponding to a period of 189.12 seconds and 23.51 seconds respectively) with quality factor Q values of 3.35 and 8.58 respectively. Figure \ref{fig:psd} (c) shows the PSD of a class $\psi$ observed by \textit{Insight}-HXMT, where the QPO frequencies are 5.76 $\times{10^{-3}}$ Hz (Q=6.15,  periods of 173.63 seconds) and 5.89$\times{10^{-2}}$ Hz (Q=6.64, periods of 16.97 seconds). For comparison, we present the PSD for a class $\kappa$ variability in Figure \ref{fig:psd} (e), and only one low-frequency QPO with frequency at 6.10 $\times10^{-3}$ Hz (Q=4.77) is found, which roughly corresponds to a period of 163.85 seconds (see also Table 2). The QPO with a period of around $10$ seconds is evident in class $\psi$ variability but with a lower Q value (e.g., $\leq2$) in most PSDs. For the observations of class $\psi$ and $\kappa$ with GTI over 1200 seconds, the PSD fitting results of QPOs are presented in Table \ref{tab:psdtable}. The signal-to-noise ratio of PSDs is low for the observations with relatively short GTIs (less than 1200 seconds), where two examples of light curves are shown in Table \ref{tab:nobID} in the appendix.

\begin{table*}
	\centering
	\caption{PSD analysis for three typical light curves observed by \textit{Insight}-HXMT and NICER. $Q$ means the quality factor of each frequency component.}
	\label{tab:psdtable}
	\begin{tabular}{ccccccccc} 
		\hline
		 Class & Obs ID & ${\rm QPO_1}$(s) & $Q_1$& $\rm rms_1$ & $\rm {QPO_2}$(s) & $Q_2$ & $\rm rms_2$ & ${\chi}^2$\\
		\hline
$\psi$&0103010107&$122.23^{+22.80}_{-12.95}$&$3.04^{+3.51}_{-1.14}$&$25.19^{+2.00}_{-2.11}$&$10.72^{+0.78}_{-0.67}$&$1.76^{+0.68}_{-0.49}$&$14.22^{+1.58}_{-1.66}$&1.31\\
$\psi$&0103010108&$146.78^{+20.78}_{-13.97}$&$1.64^{+0.85}_{-0.58}$&$29.66^{+3.07}_{-3.32}$&$14.14^{+0.68}_{-0.62}$&$3.42^{+2.84}_{-1.61}$&$10.16^{+2.64}_{-2.28}$&1.44\\
$\psi$&1103010102&$189.12^{+8.11}_{-6.90}$&$3.35^{+0.97}_{-0.86}$&$53.06^{+4.66}_{-5.43}$&$23.51^{+0.41}_{-0.47}$&$8.58^{+6.44}_{-3.23}$&$6.12^{+1.10}_{-1.21}$&1.87\\
$\psi$&P010133000103&$111.71^{+21.57}_{-16.53}$&$1.69^{+1.02}_{-0.62}$&$33.50^{+4.27}_{-4.69}$&$20.06^{+3.28}_{-2.86}$&$2.43^{+3.31}_{-1.29}$&$8.48^{+2.46}_{-2.30}$&1.28\\
$\psi$&P010133000105&$173.63^{+7.83}_{-8.07}$&$6.15^{+3.89}_{-2.18}$&$45.17^{+5.10}_{-5.67}$&$16.97^{+1.30}_{-1.19}$&$6.64^{+151}_{-3.69}$&$7.13^{+2.11}_{-3.07}$&1.40\\
$\psi$&P010133000106&$166.93^{+10.66}_{-10.04}$&$1.72^{+0.61}_{-0.46}$&$52.89^{+4.37}_{-4.88}$&$19.38^{+3.41}_{-1.72}$&$1.25^{+0.98}_{-0.66}$&$14.54^{+2.43}_{-2.29}$&1.38\\
$\psi$&P010133000107&$168.53^{+3.84}_{-4.92}$&$5.59^{+1.15}_{-0.77}$&$55.43^{+3.40}_{-3.36}$&$19.51^{+2.96}_{-1.50}$&$1.55^{+1.03}_{-0.74}$&$12.44^{+2.25}_{-2.19}$&1.79\\
$\psi$&P010133000111&$152.28^{+32.64}_{-39.37}$&$1.02^{+1.65}_{-0.45}$&$26.40^{+4.21}_{-4.39}$&$17.07^{+6.16}_{-2.85}$&$0.85^{+4.43}_{-0.47}$&$14.51^{+3.95}_{-6.60}$&1.14\\
$\psi$&P010133000112&$183.65^{+10.99}_{-27.21}$&$17.37^{+76.46}_{-14.24}$&$37.21^{+12.75}_{-20.60}$&$16.27^{+1.99}_{-1.35}$&$1.22^{+0.88}_{-0.51}$&$14.71^{+2.46}_{-2.73}$&1.04\\
$\psi$&P010133000113&$179.98^{+103.0}_{-37.54}$&$1.09^{+2.21}_{-0.70}$&$27.86^{+5.36}_{-6.03}$&$14.67^{+1.60}_{-1.36}$&$1.72^{+2.00}_{-0.85}$&$14.16^{+2.89}_{-3.28}$&0.70\\
$\psi$&P010133000114&$143.23^{+236.3}_{-26.88}$&$2.98^{+6.09}_{-2.44}$&$24.74^{+5.55}_{-7.42}$&$14.82^{+2.14}_{-1.76}$&$1.35^{+0.93}_{-0.57}$&$14.51^{+2.28}_{-2.55}$&0.75\\	$\psi$&P010131000101&$201.97^{+54.56}_{-29.60}$&$2.03^{+1.93}_{-1.30}$&$40.09^{+9.69}_{-12.37}$&$11.01^{+1.40}_{-0.82}$&$4.30^{+6.50}_{-3.13}$&$5.57^{+2.15}_{-2.04}$&1.21\\
$\psi$&P010131000102&$184.45^{+38.67}_{-36.39}$&$1.63^{+1.74}_{-0.87}$&$37.38^{+7.24}_{-9.23}$&$23.18^{+1.41}_{-1.03}$&$11.59^{+42.87}_{-7.86}$&$7.49^{+1.95}_{-2.67}$&1.04\\
$\kappa$&P010131000104&$200.92^{+42.55}_{-24.79}$&$3.03^{+3.19}_{-1.88}$&$49.19^{+11.56}_{-14.54}$&-&-&-&0.91\\
$\kappa$&1103010103&$179.61^{+43.91}_{-37.06}$&$4.89^{+12.84}_{-3.87}$&$34.05^{+9.53}_{-15.19}$&-&-&-&1.27\\
$\kappa$&1103010106&$180.68^{+283.9}_{-13.51}$&$2.53^{+2.28}_{-2.11}$&$42.42^{+10.30}_{-12.09}$&-&-&-&1.41\\
$\kappa$&1103010109&$212.14^{+75.30}_{-44.52}$&$2.91^{+6.44}_{-1.86}$&$41.41^{+12.03}_{-13.68}$&-&-&-&1.27\\
$\kappa$&1103010112&$163.85^{+16.96}_{-14.49}$&$4.77^{+69.7}_{-3.03}$&$37.05^{+10.19}_{-12.49}$&-&-&-&1.70\\
		\hline
	\end{tabular}
\end{table*}

We further adopt an independent time-varying analysis called ‘`Stepwise Filter Correlation" method (SFC) to catch the superposed signals (e.g. mini pulses) in the light curves, where this method was successfully used in searching for the superimposed periodic components in Gamma-ray bursts \citep{gao2012stepwise}. We briefly introduce this method as follows. Firstly, we use a series of low-pass filters with gradually increasing cut-off frequency $f_{i}$ ($i=1,2,..., N$) to filter the time series signals and obtain a series of corresponding filtered light curves $\rm RLC_{i}$ ($i=1,2,..., N$). Then, we perform the correlation analysis between $\rm RLC_{i}$ and $\rm RLC_{i+1}$ and get the correlation coefficients $R_{i}$. If $R_{k}$ suddenly decreases compared with $R_{k-1}$ or $R_{k+1}$ means that the original light curves signals contain the corresponding frequency component $f_{k}$. Detailed discussion and comparison with PSD analysis can be found in \cite{gao2012stepwise}. We present the results in the right panel of Figure \ref{fig:psd}, where two prominent dips can be found and the frequencies in the dips are corresponding to the QPOs as found in the PSD analysis. For comparison, there is no second dip in the $\kappa$ state (bottom panel in Figure \ref{fig:psd}). The periods derived from the SFC method are roughly consistent with those derived from PSD results, which further supports that the mini pulses are periodic in the new class $\psi$. The long periods range from 105 seconds to 180 seconds and the periods of short pulses range from 8 to 18 seconds in the $\psi$ pattern, where the periods of the long pulse are a little bit shorter than those of class $\kappa$ (115 to 220 seconds).

\begin{table*}
	\centering
	\caption{SFC analysis on the new class $\psi$ and class $\kappa$. The results of SFC analysis for class $\psi$ and  $\kappa$ are based on the observations of  \textit{Insight}-HXMT and NICER.}
	\label{tab:sfctable}
	\begin{tabular}{ccccc}
		\hline
		Variability class & Obs ID & Time segments (s) & Period 1 (s) & Period 2 (s)\\
		\hline
$\psi$&0103010105&27650-27850&132.5&7.6\\
$\psi$&0103010106&110-740&134.9&9.8\\
$\psi$&0103010107&50-720&134.7&8.2\\
$\psi$&0103010107&1330-2000&125.8&10.0\\
$\psi$&0103010108&33200-34200&125.9&11.2\\
$\psi$&1103010102&50-800&134.9&8.5\\
$\psi$&1103010102&1000-2150&117.5&9.7\\
$\psi$ &P010133000103&5500-5750&108.7&16.7\\
$\psi$ &P010133000104&1100-1400&176.6&16.7\\
$\psi$ &P010133000105&800-1200&124.8&16.7\\
$\psi$ &P010133000106&100-500&154.9&18.1\\
$\psi$ &P010133000106&1940-2140&144.5&14.8\\
$\psi$ &P010133000107&5850-6350&164.8&14.6\\
$\psi$ &P010133000110&5800-6500&108.7&12.7\\
$\psi$ &P010133000111&0-1500&176.6&15.6\\
$\psi$ &P010133000112&400-900&143.4&11.8\\
$\psi$ &P010133000113&6300-7100&116.5&14.5\\
$\psi$ &P010133000114&300-1400&164.8&13.6\\
$\psi$ &P010133000115&150-700&143.4&12.7\\
$\psi$ &P010131000101&0-350&124.8&15.6\\
$\psi$ &P010131000102&6300-6600&153.7&11.0\\
$\kappa$&P010131000103&0-1100&189.3&-\\
$\kappa$&P010131000104&0-1000&189.3&-\\
$\kappa$&P010131000105&0-1000&176.6&-\\
$\kappa$&1103010103&66500-67300&176.6&-\\
$\kappa$&1103010105&61000-61900&217.4&-\\
$\kappa$&1103010106&0-2000&202.9&-\\
$\kappa$&1103010107&22700-23400&176.6&-\\
$\kappa$&1103010108&27900-28950&202.9&-\\
$\kappa$&1103010109&5570-6460&202.0&-\\
$\kappa$&1103010112&0-1200&218.8&-\\
$\kappa$&1103010112&1300-2000&166.0&-\\
$\kappa$&1103010113&0-500&124.9&-\\
$\kappa$&1103010113&5600-6100&116.5&-\\
$\kappa$&1103010115&77700-78300&164.8&-\\
$\kappa$&1103010116&0-600&202.9&-\\
		\hline
	\end{tabular}
\end{table*}

\subsection{Spectral analysis}
\begin{figure*}
    \centering
	\includegraphics[scale=0.49]{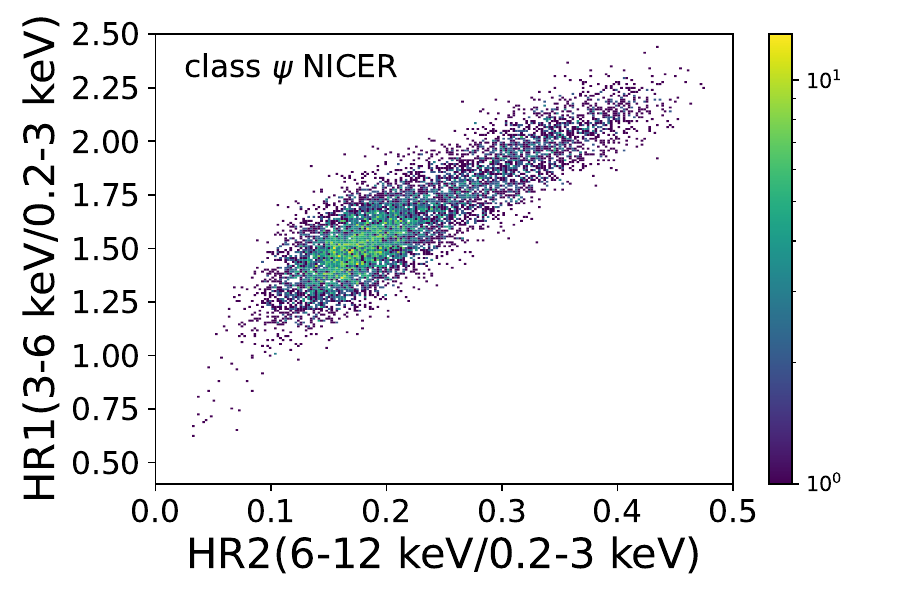}
	\includegraphics[scale=0.49]{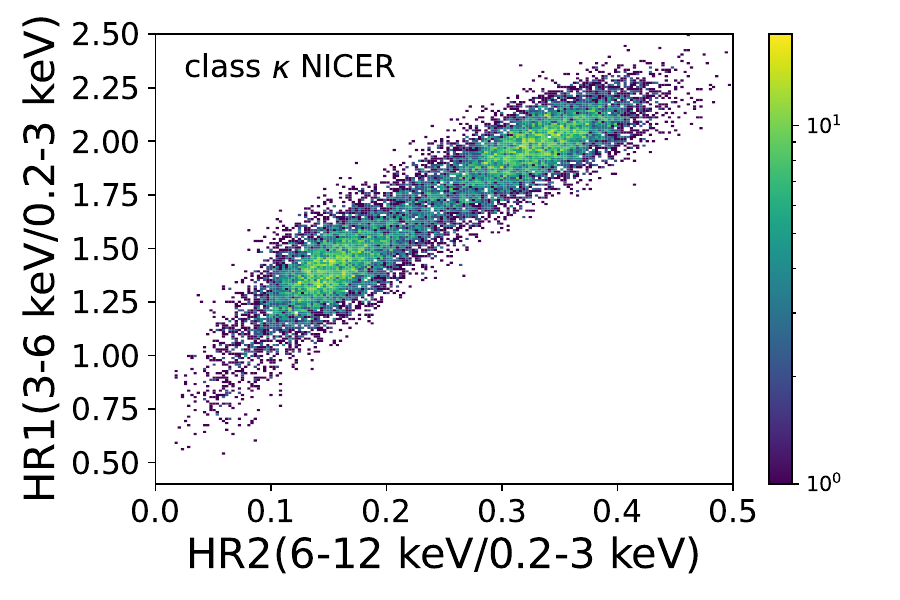}
        \includegraphics[scale=0.49]{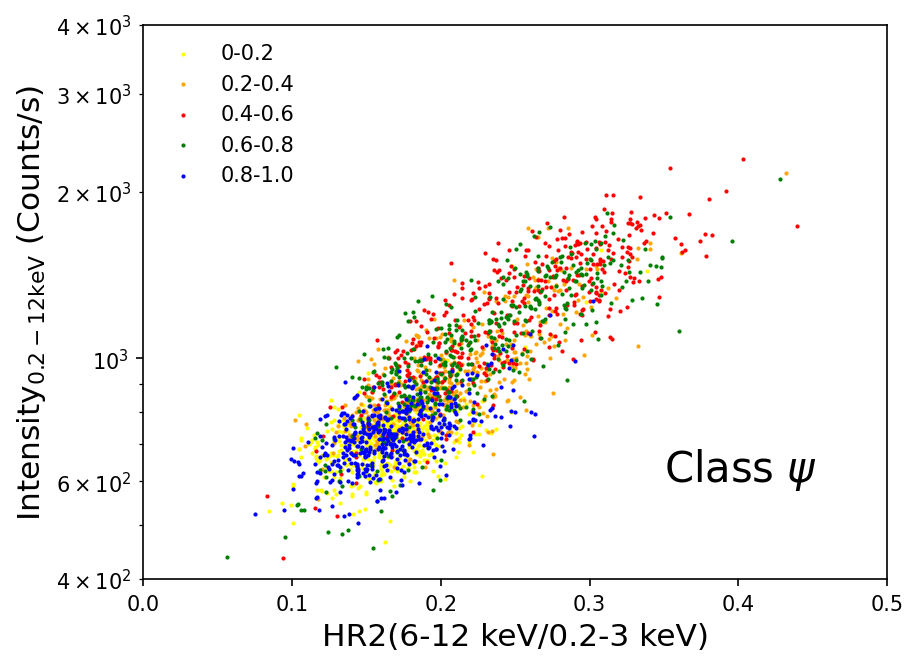}
        \includegraphics[scale=0.49]{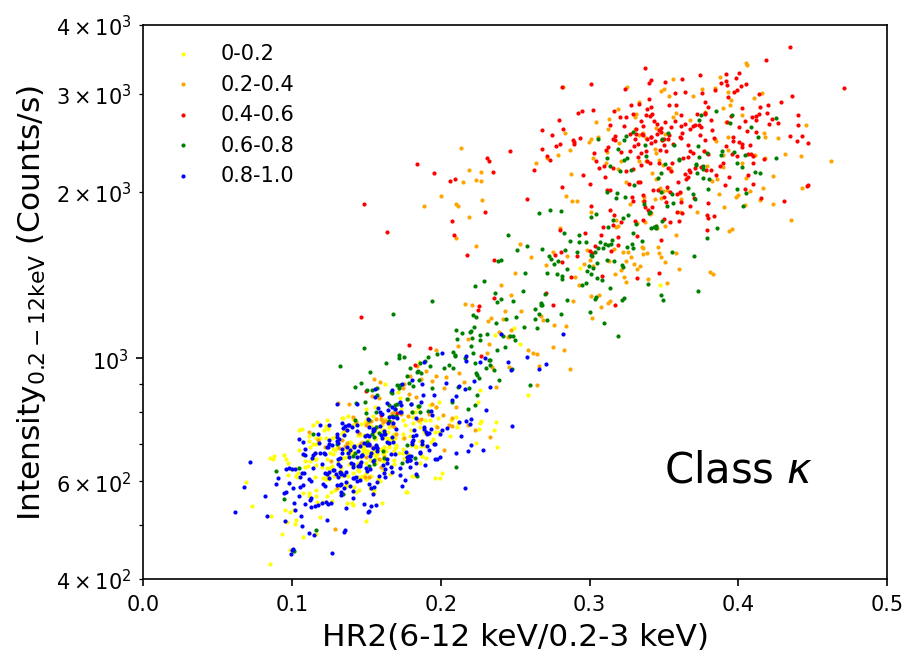}
    \caption{The CCD for 5 observations in classes $\psi$ (upper-left) and 10 observations in $\kappa$ (upper-right) as observed by NICER, where $\rm HR_{1}$=3-6 keV/0.2-3 keV, $\rm HR_{2}$=6-12 keV/0.2-3 keV and the color bar represents the dot density. The low-left cluster in the two CCDs corresponds to State C defined in \citep{belloni2000model}. The left and right panels in the bottom represent the phase-resolved HID for two observations of class $\psi$ (lower left, obsID 0103010107) and $\kappa$ (lower right, obsID 1103010102) respectively, where the different colors represent the different phase intervals.}
    \label{fig:ccd}
\end{figure*}

We present the CCD for 5 observations of class $\psi$ (MJD 57934 to 57959) and 10 for class $\kappa$ (MJD 57962 to 57974) from NICER in the left and right panel of Figure \ref{fig:ccd} respectively, where the hardness ratios are defined as $\rm HR_{1}$=(3-6 keV)/(0.2-3 keV), $\rm HR_{2}$=(6-12 keV/(0.2-3 keV). The observation IDs from these two classes are shown in Table \ref{tab:obID}. Based on the CCD results, it can be found that the class $\kappa$ shows bimodal distribution while class $\psi$ only shows one cluster of points, where the class $\psi$ corresponds to the distribution with a lower hardness ratio of class $\kappa$. The lower left clusters in the two CCDs are mainly contributed by the low-flux phase in the light curves, which corresponds to State C as defined in \citet{belloni2000model}. The mini pulses of class $\psi$ are located at the upper-right part of the CCD, identified as State A. These rapid oscillations cannot be maintained at higher luminosity for a long enough time to converge and form a cluster on the CCD. While the second cluster with the higher HR1 and HR2 in the upper right part of class $\kappa$ CCD is mainly from the high rate phase in the main pulse, which corresponds to State B. 

In order to further explore the evolution of the main pulses, we divide the folded main pulse into 5 different phase intervals (0-0.2, 0.2-0.4, 0.4-0.6, 0.6-0.8, 0.8-1.0), where the detailed implementation steps are presented in Appendix \ref{A1}. The phase-resolved hardness intensity diagram (HIDs) are plotted in Figure \ref{fig:ccd} for class $\psi$ and $\kappa$. In the HID of class $\psi$, different phase intervals are mixed since the variation between peaks and dips of mini pulses exists in each phase interval of the main pulse. Compared with the CCD results shown in Figure \ref{fig:ccd}, we propose that the extra upper right cluster of class $\kappa$ in CCD is mainly attributed to the harder intermediate phase intervals with higher intensity (0.2-0.4, 0.4-0.6, 0.6-0.8). We can also see the evolution track of main pulse: from phase 0 to 0.6, both class $\psi$ and $\kappa$ evolve towards the upper right from lower left in the HID. It becomes fainter and softer from phase 0.6 to 1.0. There is an apparent positive correlation between the count rates and hardness, which suggest that the spectrum is harder as it becomes brighter.

\begin{figure*}
    \centering
    \subfigure[]{\includegraphics[width=0.49\textwidth]{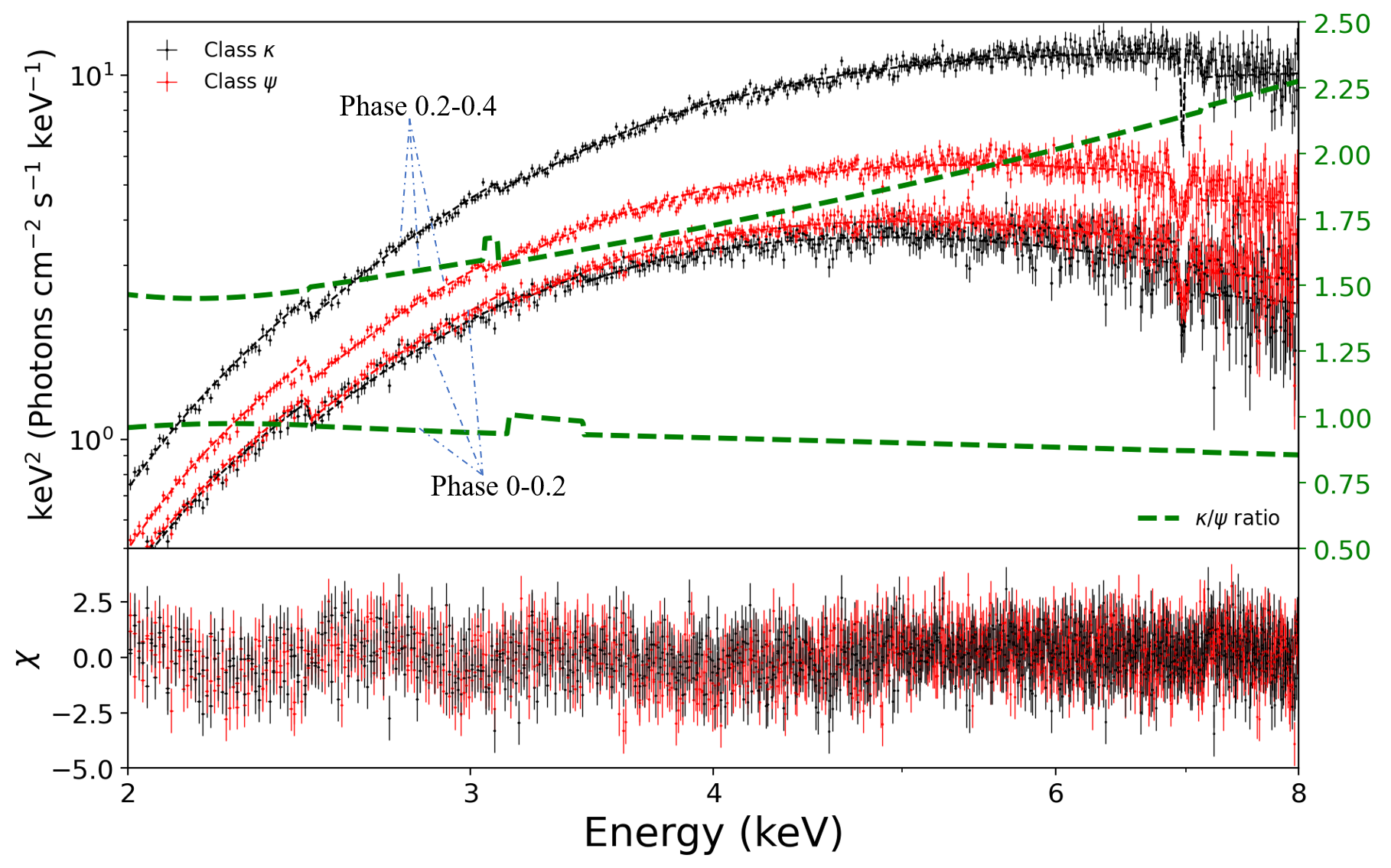}}\label{fig:speca}
    \subfigure[]{\includegraphics[width=0.49\textwidth]{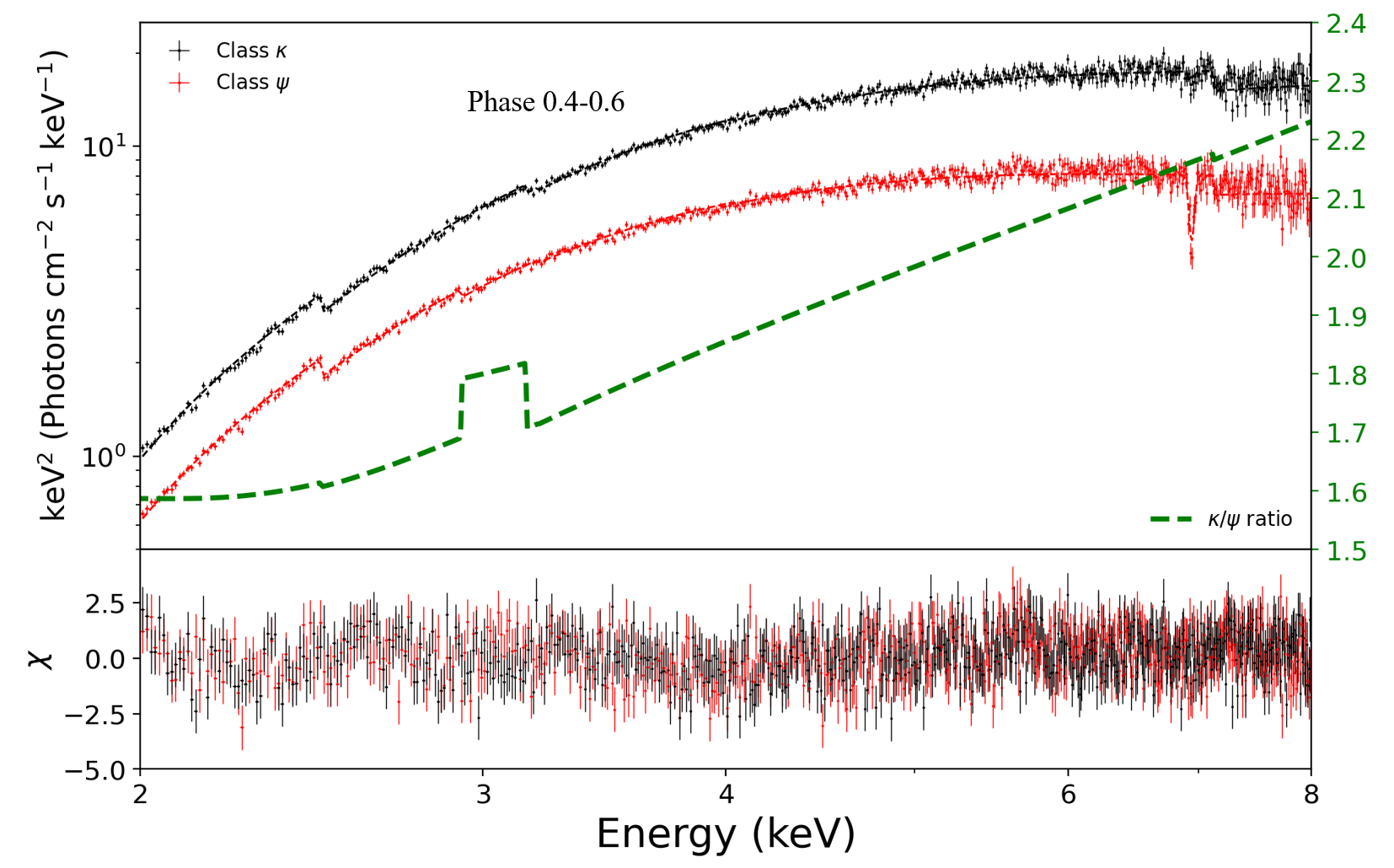}}\label{fig:specb}
    \subfigure[]{\includegraphics[width=0.49\textwidth]{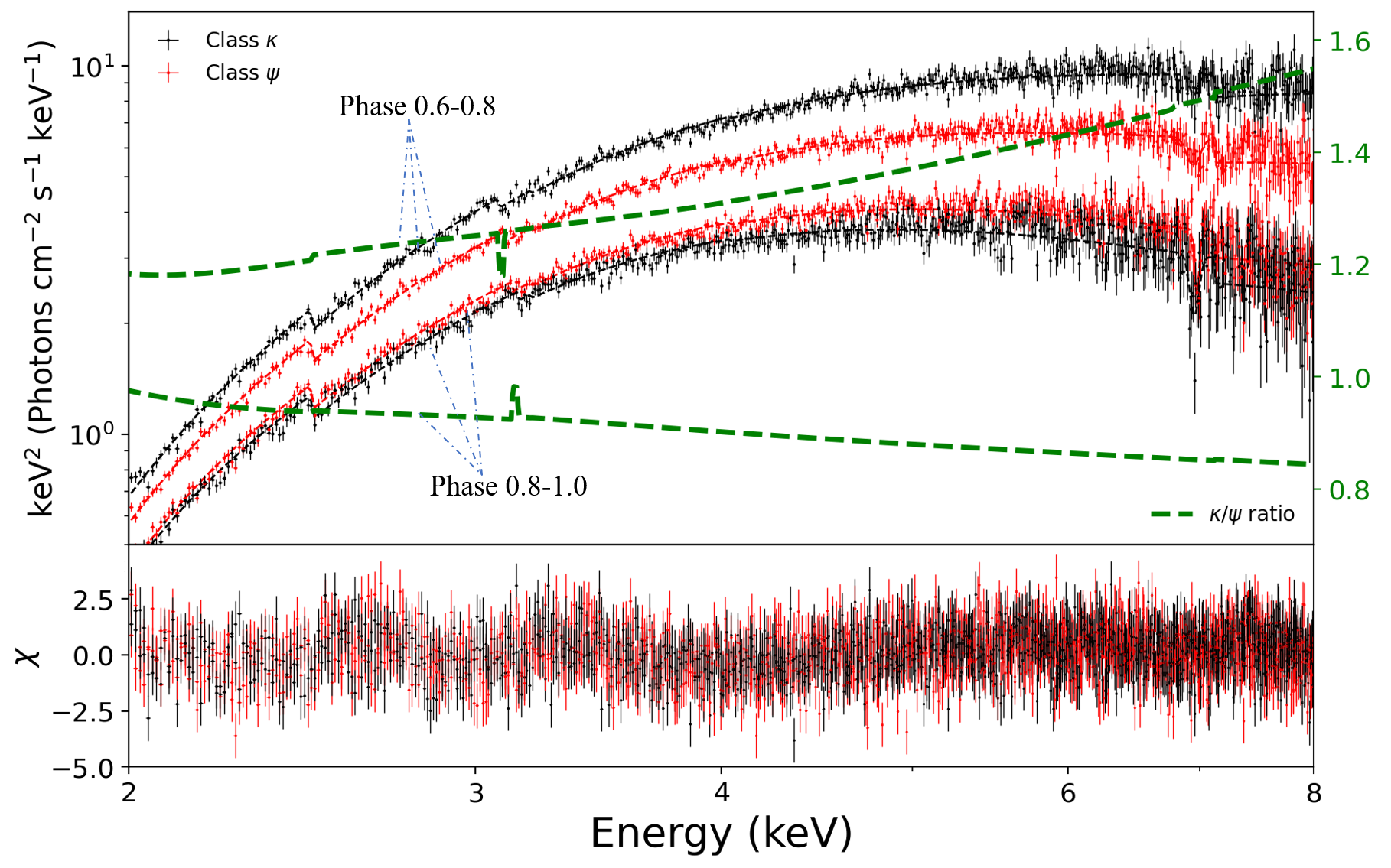}}\label{fig:specc}
    \subfigure[]{\includegraphics[width=0.49\textwidth]{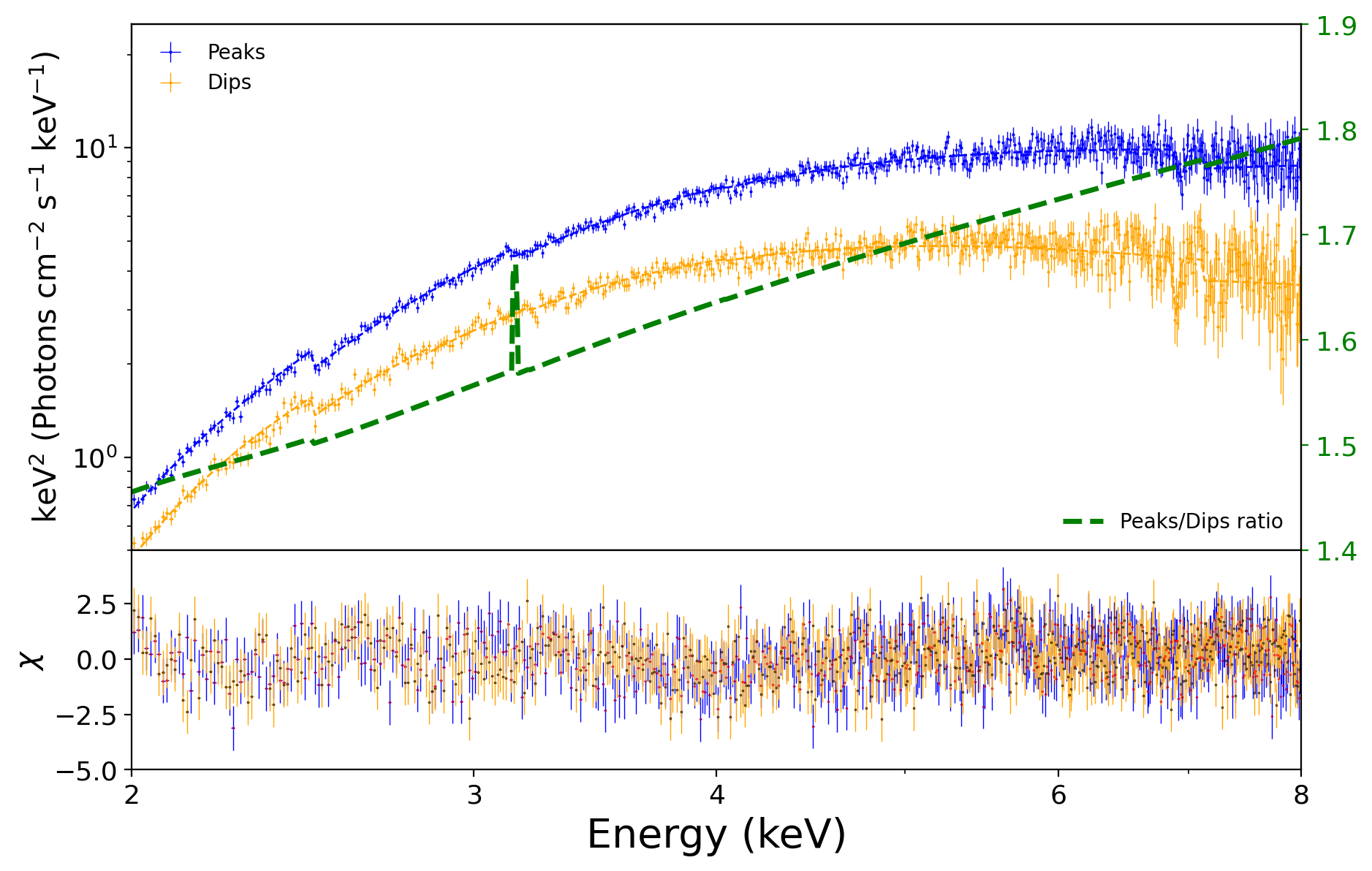}}
    \caption{The first three panels (a)-(c) represent the folded NICER 2-10 keV spectra of five different phase intervals 0-0.2, 0.2-0.4, 0.4-0.6, 0.6-0.8, 0.8-10 in class $\psi$ (red, ObsID 0103010107) and $\kappa$ (black, ObsID 1103010102). The last panel (d) are the NICER spectra of the peaks (blue) and dips (orange) of mini pulses in class $\psi$. The green dashed line shows the ratio between the two fitted models of the spectra and it removes the discrepancy brought by the fitted Gaussian line. The model we adopt is {\tt\string Tbabs*edge*gabs*simpl*diskbb}. The {\tt\string edge} is included to fit the NICER absorption edges and a Gaussian absorption line is between 6.9-7.0 keV (Fe XXVI Ly$\alpha$).}
    \label{fig:spec}
\end{figure*}

The NICER spectra of five different phase intervals in the folded main pulse for observations are shown in Figure \ref{fig:spec} for two classes respectively. We fit the spectra with model {\tt\string Tbabs*edge*gabs*simpl*diskbb}, where {\tt\string Tbabs} component is equivalent hydrogen column density in the interstellar medium for the X-ray absorption and the {\tt\string edge} is for the NICER absorption edges of transmission through windows or filters\footnote{https://heasarc.gsfc.nasa.gov/docs/nicer/analysis\_threads/arf-rmf}. The edge locates at around 2.8-3.5 keV, which corresponds to the absorption edge features of reflectivity of the gold shell at the NICER X-ray Concentrator. We include a {\tt\string gabs} model to fit the strong absorption line which is at 6.9-7.0 keV (Fe XXVI Ly$\alpha$). We adopt a {\tt \string simpl} to model the non-thermal emission and {\tt \string diskbb} for the thermal component of the disk. All the parameters of the above models are free to vary except for some insignificant Gaussian absorption lines that are fixed in some phase intervals, where the best-fitting results are presented in Table \ref{tab:spectable}. 

\begin{figure}
    \centering
    \includegraphics[width=0.4\textwidth]{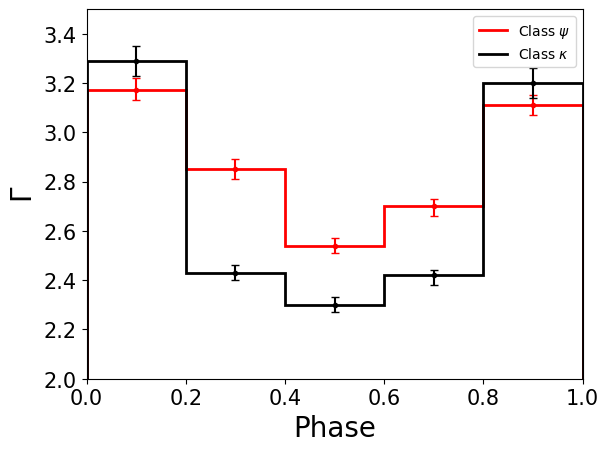}
    \includegraphics[width=0.4\textwidth]{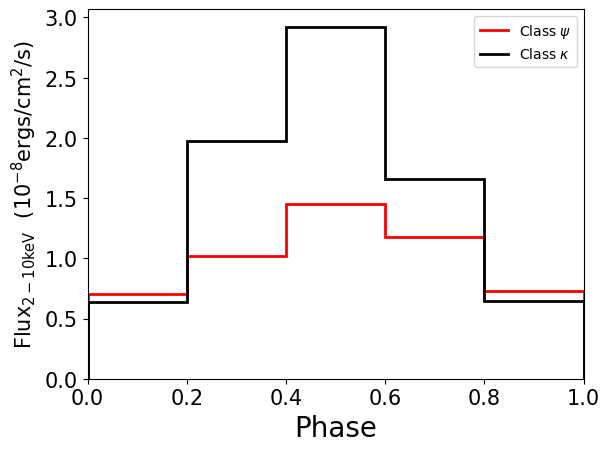}
    \caption{The evolved photon index of {\tt\string simpl} and flux in different phase intervals. 
    }
    \label{fig:fitpar}
\end{figure}

The spectra are shown in Figure \ref{fig:spec} and parameters evolution in the folded main pulses are in Figure \ref{fig:fitpar}. The ratio between two fitted models of class $\kappa$ and $\psi$ spectrum is depicted in green dashed lines and the discrepancy brought by the fitted Gaussian line is removed. As shown in Figure \ref{fig:spec}(a) and \ref{fig:spec}(c), there is no significant difference in the relatively quiet phase (phase interval of 0-0.2 and 0.8-1.0) between the class $\psi$ and $\kappa$. The ratio between the two classes stays steadily declining from 2 to 10 keV in these two phase intervals and class $\psi$ spectrum is a little harder than $\kappa$. From phase 0 to 0.6, the photon index of class $\psi$ varies from $3.17^{+0.05}_{-0.04}$ to $2.54\pm{0.03}$ while from $3.28\pm{0.06}$ to $2.29\pm{0.03}$ in class $\kappa$. The main differences in spectra between the two classes are around the intermediate phases, particularly around 0.4–0.6, class $\kappa$ has a harder spectrum and higher rate than class $\psi$, which is consistent with that there is an additional upper-right cluster of class $\kappa$ in Figure \ref{fig:ccd}, and class $\psi$ changes into class $\kappa$ as the count rates increase, shown in Figure \ref{fig:lc1}. 
To further investigate the spectral variations of mini pulses in the class $\psi$, we select the local maximum and minimum count rate values of each mini pulse in the light curves and extract the spectra in 2s duration for each peak/dip phase of mini pulses (see Figure \ref{fig:peaks}). The folded spectra are shown in Figure \ref{fig:spec}(d) and the best-fitting parameters are listed in Table \ref{tab:minitable}. The peak phase spectrum is harder with a photon index of $2.41\pm{0.04}$ compared with $2.99\pm{0.06}$ in the dip phase spectrum. This result is also consistent with that when the source is brighter, the spectrum is harder. The best-fitting temperatures for the disk are roughly constant across all the folded phase spectra, falling between 0.18 and 0.20 keV.

\begin{table}
        \centering
	\caption{The fitting parameters of peaks and dips of mini pulses in class $\psi$. The error of parameters is determined to be within the 90\% confidence region. }
	\label{tab:minitable}
        \begin{tabular*}{0.35\textwidth}{ccc}
        \toprule
        Parameters & Peaks & Dips\\
        \midrule
        ${N_{\rm H}}$ ($\times 10^{22}$) & $9.08\pm{0.28}$ & $9.80\pm{0.51}$\\
        
        ${E_{\rm edge}}$ (keV) & $3.14\pm{0.05}$ & $3.12^{+0.20}_{-0.07}$\\
        
        ${\tau_{\rm edge}}$ ($\times 10^{-2}$) & $6.21^{+1.87}_{-1.71}$ & $4.48^{+2.25}_{-2.94}$\\

        $E_{\rm Fe}$ (keV) & $6.93\pm{0.05}$ & $6.90^{+0.05}_{-0.04}$\\
        
        $\sigma_{\rm Fe}$ ($\times 10^{-2})$ keV) & $2.89^*$ & $2.00^*$\\
        
        Strength ($\times 10^{-2}$) & $1.54\pm{0.55}$ & $2.12^{+1.67}_{-1.36}$\\
        
        $\Gamma$ & $2.41\pm{0.04}$ & $2.99\pm{0.06}$\\

        $f_{\rm scat}$ ($\times 10^{-2}$) & $1.41^{+0.55}_{-0.54}$ & $1.71^{+0.60}_{-0.49}$\\

        $\rm T_{in}$ & $0.19\pm{0.01}$ & $0.19\pm{0.01}$\\
        
        Norm ($\times 10^{8}$) & $0.99^{+1.28}_{-0.45}$ & $1.10^{+1.08}_{-0.51}$\\
        
        ${\chi}^2/$d.o.f & $749/721$ & $822/720$\\
        \bottomrule
	\end{tabular*}
\end{table}

\begin{table*}
	\centering
	\caption{The fitting parameters of folded spectra in 5 different phase intervals in class $\psi$ (ObsID 0103010107) and $\kappa$ (ObsID 1103010102). The model for each folded spectrum is {\tt\string Tbabs*edge*gabs*simpl*diskbb}. The {\tt\string edge} is for NICER absorption edge. The error of parameters is determined to be within the $90\%$ confidence region. In some phase intervals, Gaussian absorption lines are not apparent and these parameters are fixed without given error bars to maintain consistency.}
	\label{tab:spectable}
        \setlength\tabcolsep{3pt}
	\begin{tabular}{ccccccccccc}
		\hline
        Phase & \multicolumn{2}{c}{0 - 0.2} & \multicolumn{2}{c}{0.2 - 0.4} & \multicolumn{2}{c}{0.4 - 0.6} & \multicolumn{2}{c}{0.6 - 0.8} & \multicolumn{2}{c}{0.8 - 1.0}\\
        Class & $\psi$ & $\kappa$ & $\psi$ & $\kappa$ & $\psi$ & $\kappa$ & $\psi$ & $\kappa$ & $\psi$ & $\kappa$  \\
        \hline
        \\
        ${N_{\rm H}}$ ($\times 10^{22}$) 
        & $10.47\pm{0.29}$ & $10.68^{+0.46}_{-0.41}$ & $9.93^{+0.27}_{-0.18}$ & $9.48\pm{0.23}$ & $9.02^{+0.27}_{-0.24}$
        & $9.29^{+0.20}_{-0.19}$ & $9.43\pm{0.23}$ & $8.94^{+0.23}_{-0.25}$ & $10.27^{+0.29}_{-0.27}$ & $10.55^{+0.43}_{-0.41}$
        \\
        ${E_{\rm edge}}$ (keV)
        & $3.13\pm{0.04}$ & $3.43^{+0.16}_{-0.10}$ & $3.04_{-0.13}^{+0.05}$ & $3.09\pm{0.04}$ & $2.92\pm{0.04}$
        &$3.15\pm{0.04}$ &$3.10\pm{0.03}$ & $3.08\pm{0.03}$ & $3.13\pm{0.05}$ & $3.39^{+0.07}_{-0.09}$
        \\
        ${\tau_{\rm edge}}(\times 10^{-2})$
        & $7.49^{+1.83}_{-1.75}$ & $5.09^{+1.98}_{-2.07}$ & $4.94_{-1.79}^{+1.75}$ & $6.51\pm{1.44}$ &$5.77^{+2.44}_{-1.73}$
        & $6.45^{+1.08}_{-1.17}$ & $6.48^{+1.44}_{-1.41}$ & $6.40^{+1.70}_{-1.47}$ & $6.15^{+1.85}_{-1.63}$ & $6.36^{+2.05}_{-2.03}$
        \\
        $E_{\rm Fe}$ (keV)
        & $6.97\pm{0.03}$ & $6.97\pm{0.02}$ & $6.94\pm{0.02}$ & $6.97\pm{0.04}$ & $6.94\pm{0.02}$
        &$6.95^*$ & $6.94^{+0.04}_{-0.05}$ & $6.95\pm{0.08}$ & $6.97^{+0.04}_{-0.03}$ & $6.95\pm{0.04}$
        \\
        $\sigma_{\rm Fe}$ ($\times 10^{-2}$ keV)
        & $4.14\pm{3.95}$ & $ 1.26\pm{0.16}$ & $3.24^{+2.54}_{-1.19}$ & $1.50^*$ & $1.97^{+4.36}_{-0.50}$
        & $2.00^*$ & $5.41^{+3.48}_{-5.30}$ & $3.40_{-3.10}^{+2.07}$ & $2.13^*$ & $3.91^*$
        \\
        Strength ($\times 10^{-2}$)
        & $3.90\pm{0.72}$  & $2.59\pm{0.14}$  & $2.90\pm{0.54}$ &$1.60^{+1.00}_{-0.79}$ &$2.71^*$
        &$2.00^*$ & $2.33^{+1.14}_{-0.85}$ & $1.08^{+0.98}_{-0.72}$ & $1.85^*$ & $3.68\pm{0.92}$
        \\
        $\Gamma$
        & $3.17^{+0.05}_{-0.04}$ & $3.28\pm{0.06}$ & $2.85\pm{0.04}$ & $2.43\pm{0.03}$ &$2.54\pm{0.03}$
        &$2.29\pm{0.03}$ & $2.70^{+0.03}_{-0.04}$ & $2.42^{+0.02}_{-0.04}$ & $3.11\pm{0.04}$ & $3.24\pm{0.06}$
        \\
        $f_{\rm scat}$ ($\times 10^{-2}$)
        & $0.95^{+0.24}_{-0.22}$ & $1.18^{+0.36}_{-0.26}$  & $1.66_{-0.34}^{+0.36}$ & $1.31^{+0.35}_{-0.33}$ & $2.06^{+0.63}_{-0.67}$
        & $1.17^{+0.35}_{-0.28}$ & $1.29^{+0.35}_{-0.32}$ & $1.38^{+0.50}_{-0.52}$ & $1.27^{+0.27}_{-0.30}$ & $1.18^{+0.32}_{-0.25}$
        \\
        $\rm T_{in}$
        & $0.18\pm{0.007}$ & $0.19\pm{0.007}$  & $0.20\pm{0.009}$ & $0.20\pm{0.009}$ &$0.20^{+0.007}_{-0.005}$
        & $0.19\pm{0.009}$ & $0.19^{+0.008}_{-0.009}$ & $0.19\pm{0.01}$ & $0.19^{+0.006}_{-0.008}$ & $0.18\pm{0.007}$
        \\
        Norm ($\times 10^{8}$) 
        & $3.67^{+2.38}_{-1.34}$ & $2.74^{+1.75}_{-1.73}$ & $0.99_{-0.33}^{+0.57}$ & $1.20^{+0.83}_{-0.43}$ &$0.62^{+0.40}_{-0.30}$
        & $1.66^{+0.99}_{-0.62}$ & $1.38^{+0.96}_{-0.50}$ & $1.02_{-0.44}^{+1.28}$ & $2.01^{+1.39}_{-0.67}$ & $2.74^{+1.83}_{-1.10}$
        \\
        ${\chi}^2/$d.o.f
        & $872/719$ & $811/719$ & $898/718$ & $750/720$  &$903/720$ & $881/722$ & $880/719$ & $849/719$ & $933/721$ & $824/721$\\
		\hline
	\end{tabular}
\end{table*}

\section{Conclusion and Discussion}

Based on the X-ray light curves and color-color diagrams of GRS 1915+105 observed by NICER and \textit{Insight}-HXMT, we discover a new pattern of variability, which we name as $\psi$. The new class $\psi$ is mainly found in the rise stage of the recent long outburst of GRS 1915+105 (from MJD 57953 to 58657). The main feature is that there are several periodic mini pulses superposed on another longer periodic pulse, where the periods are $\sim10$ seconds and $\sim130$ seconds respectively based on the PSD and SFC analysis. We find that the longer period of class $\psi$ is more or less similar to that of class $\kappa$, where the mini pulses may be suppressed as increasing luminosity. The CCD of class $\psi$ shows one distribution while class $\kappa$ shows bimodal clusters.

With the launch of the Rossi X-ray Timing satellite (RXTE), the XRBs were found to show variability properties in different timescales. GRS 1915+105 shows peculiar and unusual levels of variability from milliseconds to weeks, which has been widely explored in the last several decades. Based on the light curves and color characteristics, the variability of GRS 1915+105 is classified into 14 classes \citep{belloni2000model,hannikainen2005characterizing,naik2002fast}. From the analysis of NICER and \textit {Insight}-HXMT, we find a new class, $\psi$, that is different from the former reported light curves, where the regular mini pulses are superposed on one main pulse. This new class appears in the rise stage of the long outburst from 2017 to 2021 and it changes into class $\kappa$ as flux further increases. We analyze and compare class $\psi$ with class $\kappa$ of light curves, PSD, and SFC analysis results in time-varying aspect, and also phase-resolved CCD, HID, and spectra in spectral aspect. The structure of light curves in class $\kappa$ is characterized by one longer pulse with periods of $\sim100-200$ seconds. Some light curves show a mixture of class $\psi$ and class $\kappa$ (e.g., panels i and j in Figure \ref{fig:lc2}), and the mini pulses are suppressed in class $\kappa$. In the CCD, class $\psi$ and $\kappa$ are also different, where class $\kappa$ has two clusters of points while $\psi$ has only one cluster. The additional cluster as found in class $\kappa$ may be caused by maintaining at the higher intensity state compared to class $\psi$.

No discernible difference is apparent between class $\psi$ and class $\kappa$ in the relatively quiet phase intervals. In terms of the overall quasi-period of the main pulses, the variations of different phases of class $\kappa$ are more severe in both total photon count rates in light curves and hardness of spectra compared with $\psi$. Thus, the difference becomes noticeable in the intermediate phases, the reversely harder spectrum of class $\kappa$ matches the second cluster in the top right corner of CCD results. Since the peaks of mini pulses are harder and brighter than dips in the spectra, the fluctuation between the peaks and dips of mini pulses occurring in intermediate phases makes class $\psi$ softer and fainter. The spectra of both two classes are harder when brighter during the evolution in different phases. Combined with the count rates in light curves, it may imply that a relatively higher accretion rate supports a relatively more stable inner region of the disk during the main pulses of class $\kappa$. Therefore, class $\kappa$ could maintain high count rates in the main pulses instead of suffering mini-oscillations as in class $\psi$. As a result, more seed photons from the disk are scattered into non-thermal emission in class $\kappa$ than in class $\psi$, as indicated by a larger hardness ratio and total flux ranging from 2 to 10 keV in class $\kappa$. 


The dramatic variability as found in GRS 1915+105 provides valuable clues on the disk transition and evolution. \cite{belloni2000model} proposed that, as suggested by \cite{markwardt1999variable} for a single observation, the spectra of GRS 1915+105 can be explained by the possible disk transition,  where the soft states with the full cold disk are observable and hard state without the inner cold disk. The new class $\psi$ shows strong variabilities at the timescales of $\sim 10$ seconds and $\sim 130$ seconds. Combined Figure \ref{fig:lc1} with Figure \ref{fig:lc2}, it is interesting to note that, the structure of light curves changes from class $\psi$ into class $\kappa$ and the mini pulses are suppressed as the luminosity increases. In some observations, the light curves also show a mixture of class $\kappa$ and $\psi$, where the mini pulses disappear when the flux is larger than a critical flux (see the red dot line in Figure \ref{fig:lc2}). It suggests that the matter supply in the inner region of the accretion disk may play a key role in the presence or absence of mini pulses. The mini pulses may correspond to the quick appearance and disappearance of the inner cold disk at a critical accretion rate, and the stable disk will be formed when the accretion rate is larger than this critical rate. By comparing class $\psi$ with $\kappa$ in the overall variation during the main pulses, we think the observation of this new class reveals a kind of oscillation, which is invisible in the previous classes. It is possible that even in other classes of GRS 1915+105, like $\lambda$, when the bursts in the oscillation don't last that long, tiny pulses like those in class $\psi$ may arise. Therefore, the class $\psi$ and its relation to class $\kappa$ provide us a useful clue to explore the class transition in GRS 1915+105, the appearance of $\sim$10s LFQPO and the formation of a full cold disk before the source changes into the soft state.


\section*{Acknowledgements}

QW was supported by the National SKA Program of China (2022SKA0120101), the National Natural Science Foundation of China (grants U1931203 and  12233007), and the science research grants from the China Manned Space Project with NO. CMS-CSST-2021-A06; ZY was supported in part by the Natural Science Foundation of China (grants U1838203 and U1938114), the Youth Innovation Promotion Association of CAS, and funds for key programs of Shanghai astronomical observatory.

This work made use of the publicly available data and software from the \textit{Insight}-HXMT and NICER missions. The \textit{Insight}-HXMT project is funded by the China National Space Administration (CNSA) and the Chinese Academy of Sciences (CAS). The NICER data and software are provided by the High Energy Astrophysics Science Archive Research Center (HEASARC) and NASA’s Astrophysics Data System Bibliographic Services.

\section*{Data Availability}
The data of \textit{Insight}-HXMT observations used in this paper is from the Institute of High Energy Physics Chinese Academy of Sciences (IHEPCAS) and is publicly available for download from the \textit{Insight}-HXMT website. The data of NICER observations is publicly available through NASA/HEASARC archives.

\bibliographystyle{mnras}
\bibliography{ref} 

\begin{thebibliography}{}
\makeatletter
\relax
\def\mn@urlcharsother{\let\do\@makeother \do\$\do\&\do\#\do\^\do\_\do\%\do\~}
\def\mn@doi{\begingroup\mn@urlcharsother \@ifnextchar [ {\mn@doi@}
  {\mn@doi@[]}}
\def\mn@doi@[#1]#2{\def\@tempa{#1}\ifx\@tempa\@empty \href
  {http://dx.doi.org/#2} {doi:#2}\else \href {http://dx.doi.org/#2} {#1}\fi
  \endgroup}
\def\mn@eprint#1#2{\mn@eprint@#1:#2::\@nil}
\def\mn@eprint@arXiv#1{\href {http://arxiv.org/abs/#1} {{\tt arXiv:#1}}}
\def\mn@eprint@dblp#1{\href {http://dblp.uni-trier.de/rec/bibtex/#1.xml}
  {dblp:#1}}
\def\mn@eprint@#1:#2:#3:#4\@nil{\def\@tempa {#1}\def\@tempb {#2}\def\@tempc
  {#3}\ifx \@tempc \@empty \let \@tempc \@tempb \let \@tempb \@tempa \fi \ifx
  \@tempb \@empty \def\@tempb {arXiv}\fi \@ifundefined
  {mn@eprint@\@tempb}{\@tempb:\@tempc}{\expandafter \expandafter \csname
  mn@eprint@\@tempb\endcsname \expandafter{\@tempc}}}

\bibitem[\protect\citeauthoryear{Belloni, Mendez, King, van~der Klis  \&
  Van~Paradijs}{Belloni et~al.}{1997a}]{belloni1997unstable}
Belloni T.,  Mendez M.,  King A.,  van~der Klis M.,   Van~Paradijs J.,  1997a,
  The Astrophysical Journal Letters, 479, L145

\bibitem[\protect\citeauthoryear{Belloni, Mendez, King, van~der Klis  \&
  Van~Paradijs}{Belloni et~al.}{1997b}]{belloni1997unified}
Belloni T.,  Mendez M.,  King A.,  van~der Klis M.,   Van~Paradijs J.,  1997b,
  The Astrophysical Journal Letters, 488, L109

\bibitem[\protect\citeauthoryear{Belloni, Klein~Wolt, M{\'e}ndez, van~der Klis
  \& van Paradijs}{Belloni et~al.}{2000}]{belloni2000model}
Belloni T.,  Klein~Wolt M.,  M{\'e}ndez R.,  van~der Klis M.,   van Paradijs
  J.,  2000, Astronomy \& Astrophysics, 355, 271

\bibitem[\protect\citeauthoryear{Cao et~al.,}{Cao et~al.}{2020}]{cao2020medium}
Cao X.,  et~al., 2020, SCIENCE CHINA Physics, Mechanics \& Astronomy, 63, 1

\bibitem[\protect\citeauthoryear{Capitanio, Del~Santo, Bozzo, Ferrigno,
  De~Cesare  \& Paizis}{Capitanio et~al.}{2012}]{capitanio2012peculiar}
Capitanio F.,  Del~Santo M.,  Bozzo E.,  Ferrigno C.,  De~Cesare G.,   Paizis
  A.,  2012, Monthly Notices of the Royal Astronomical Society, 422, 3130

\bibitem[\protect\citeauthoryear{Castro-Tirado, Brandt, Lund, Lapshov, Sunyaev,
  Shlyapnikov, Guziy  \& Pavlenko}{Castro-Tirado
  et~al.}{1994a}]{castro1994discovery}
Castro-Tirado A.~J.,  Brandt S.,  Lund N.,  Lapshov I.,  Sunyaev R.~A.,
  Shlyapnikov A.~A.,  Guziy S.,   Pavlenko E.~P.,  1994a, The Astrophysical
  Journal Supplement Series, 92, 469

\bibitem[\protect\citeauthoryear{Castro-Tirado, Brandt, Lund, Lapshov  \&
  Pavlenko}{Castro-Tirado et~al.}{1994b}]{1994Discovery}
Castro-Tirado A.~J.,  Brandt S.,  Lund N.,  Lapshov I.,   Pavlenko E.~P.,
  1994b, Astrophysical Journal Supplement, 92, 469

\bibitem[\protect\citeauthoryear{Chen et~al.,}{Chen et~al.}{2020}]{chen2020low}
Chen Y.,  et~al., 2020, SCIENCE CHINA Physics, Mechanics \& Astronomy, 63,
  249505

\bibitem[\protect\citeauthoryear{Fender \& Belloni}{Fender \&
  Belloni}{2004}]{fender2004grs}
Fender R.,  Belloni T.,  2004, Annu. Rev. Astron. Astrophys., 42, 317

\bibitem[\protect\citeauthoryear{Gao, Zhang  \& Zhang}{Gao
  et~al.}{2012}]{gao2012stepwise}
Gao H.,  Zhang B.-B.,   Zhang B.,  2012, The Astrophysical Journal, 748, 134

\bibitem[\protect\citeauthoryear{Guo et~al.,}{Guo
  et~al.}{2020}]{guo2020background}
Guo C.-C.,  et~al., 2020, Journal of High Energy Astrophysics, 27, 44

\bibitem[\protect\citeauthoryear{Hannikainen et~al.,}{Hannikainen
  et~al.}{2005}]{hannikainen2005characterizing}
Hannikainen D.~C.,  et~al., 2005, Astronomy \& Astrophysics, 435, 995

\bibitem[\protect\citeauthoryear{Ingram \& Done}{Ingram \&
  Done}{2012}]{ingram2012modelling}
Ingram A.,  Done C.,  2012, Monthly Notices of the Royal Astronomical Society,
  419, 2369

\bibitem[\protect\citeauthoryear{Liao et~al.,}{Liao
  et~al.}{2020a}]{liao2020background2}
Liao J.-Y.,  et~al., 2020a, Journal of High Energy Astrophysics, 27, 14

\bibitem[\protect\citeauthoryear{Liao et~al.,}{Liao
  et~al.}{2020b}]{liao2020background1}
Liao J.-Y.,  et~al., 2020b, Journal of High Energy Astrophysics, 27, 24

\bibitem[\protect\citeauthoryear{Liu et~al.,}{Liu et~al.}{2020}]{liu2020high}
Liu C.,  et~al., 2020, SCIENCE CHINA Physics, Mechanics \& Astronomy, 63, 1

\bibitem[\protect\citeauthoryear{Markwardt, Swank  \& Taam}{Markwardt
  et~al.}{1999}]{markwardt1999variable}
Markwardt C.~B.,  Swank J.~H.,   Taam R.~E.,  1999, The Astrophysical Journal
  Letters, 513, L37

\bibitem[\protect\citeauthoryear{Massaro, Ventura, Massa, Feroci, Mineo,
  Cusumano, Casella  \& Belloni}{Massaro et~al.}{2010}]{massaro2010complex}
Massaro E.,  Ventura G.,  Massa F.,  Feroci M.,  Mineo T.,  Cusumano G.,
  Casella P.,   Belloni T.,  2010, Astronomy \& Astrophysics, 513, A21

\bibitem[\protect\citeauthoryear{Massaro, Capitanio, Feroci, Mineo, Ardito  \&
  Ricciardi}{Massaro et~al.}{2020}]{massaro2020non}
Massaro E.,  Capitanio F.,  Feroci M.,  Mineo T.,  Ardito A.,   Ricciardi P.,
  2020, Monthly Notices of the Royal Astronomical Society, 495, 1110

\bibitem[\protect\citeauthoryear{Mirabel \& Rodriguez}{Mirabel \&
  Rodriguez}{1994}]{mirabel1994superluminal}
Mirabel I.,  Rodriguez L.,  1994, Nature, 371, 46

\bibitem[\protect\citeauthoryear{Motta, Williams, Fender, Titterington, Green
  \& Perrott}{Motta et~al.}{2019}]{motta2019ami}
Motta S.,  Williams D.,  Fender R.,  Titterington D.,  Green D.,   Perrott Y.,
  2019, The Astronomer's Telegram, 12773, 1

\bibitem[\protect\citeauthoryear{Naik, Rao  \& Chakrabarti}{Naik
  et~al.}{2002}]{naik2002fast}
Naik S.,  Rao A.,   Chakrabarti S.~K.,  2002, Journal of Astrophysics and
  Astronomy, 23, 213

\bibitem[\protect\citeauthoryear{Negoro et~al.,}{Negoro
  et~al.}{2018}]{negoro2018maxi}
Negoro H.,  et~al., 2018, The Astronomer's Telegram, 11828, 1

\bibitem[\protect\citeauthoryear{Reid, McClintock, Steiner, Steeghs, Remillard,
  Dhawan  \& Narayan}{Reid et~al.}{2014}]{reid2014parallax}
Reid M.,  McClintock J.,  Steiner J.,  Steeghs D.,  Remillard R.,  Dhawan V.,
  Narayan R.,  2014, The Astrophysical Journal, 796, 2

\bibitem[\protect\citeauthoryear{Remillard \& McClintock}{Remillard \&
  McClintock}{2006}]{remillard2006x}
Remillard R.~A.,  McClintock J.~E.,  2006, Annu. Rev. Astron. Astrophys., 44,
  49

\bibitem[\protect\citeauthoryear{Remillard, Morgan, McClintock, Bailyn, Orosz
  \& Greiner}{Remillard et~al.}{1997}]{remillard1997multifrequency}
Remillard R.~A.,  Morgan E.~H.,  McClintock J.~E.,  Bailyn C.~D.,  Orosz J.~A.,
    Greiner J.,  1997, arXiv preprint astro-ph/9705064

\bibitem[\protect\citeauthoryear{Shakura \& Sunyaev}{Shakura \&
  Sunyaev}{1973}]{shakura1973reprint}
Shakura N.,  Sunyaev R.,  1973, Astronomy and Astrophysics, 500, 33

\bibitem[\protect\citeauthoryear{Tanaka \& Shibazaki}{Tanaka \&
  Shibazaki}{1996}]{tanaka1996x}
Tanaka Y.,  Shibazaki N.,  1996, Annual Review of Astronomy and Astrophysics,
  34, 607

\bibitem[\protect\citeauthoryear{Yuan \& Narayan}{Yuan \&
  Narayan}{2014}]{yuan2014hot}
Yuan F.,  Narayan R.,  2014, Annual Review of Astronomy and Astrophysics, 52,
  529

\bibitem[\protect\citeauthoryear{Zhang et~al.,}{Zhang
  et~al.}{2020}]{zhang2020overview}
Zhang S.-N.,  et~al., 2020, SCIENCE CHINA Physics, Mechanics \& Astronomy, 63,
  1

\bibitem[\protect\citeauthoryear{Zoghbi et~al.,}{Zoghbi
  et~al.}{2016}]{zoghbi2016disk}
Zoghbi A.,  et~al., 2016, The Astrophysical Journal, 833, 165

\makeatother
\end{thebibliography}

\appendix

\section {Phase-resolved folded spectral analysis}
\label{A1}

\begin{figure*}
    \centering
	\includegraphics[scale=0.49]{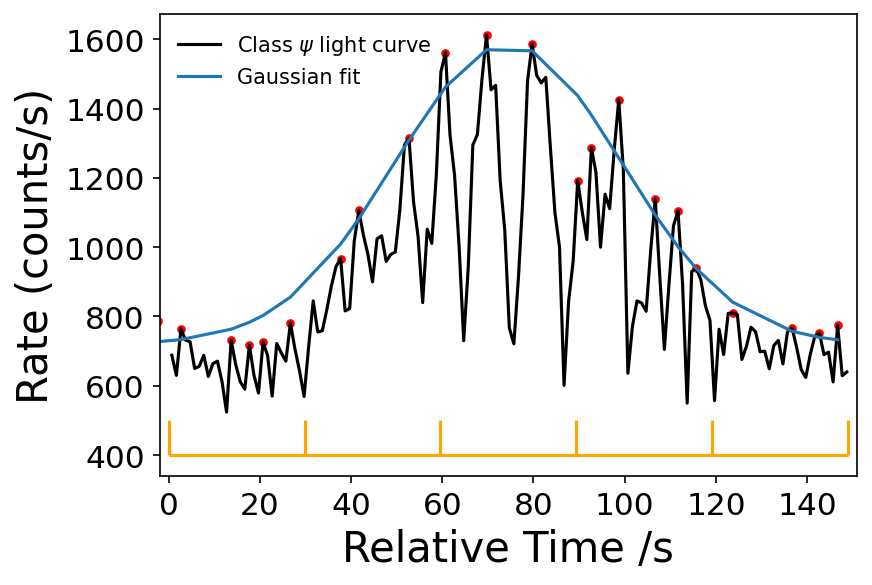}
	\includegraphics[scale=0.49]{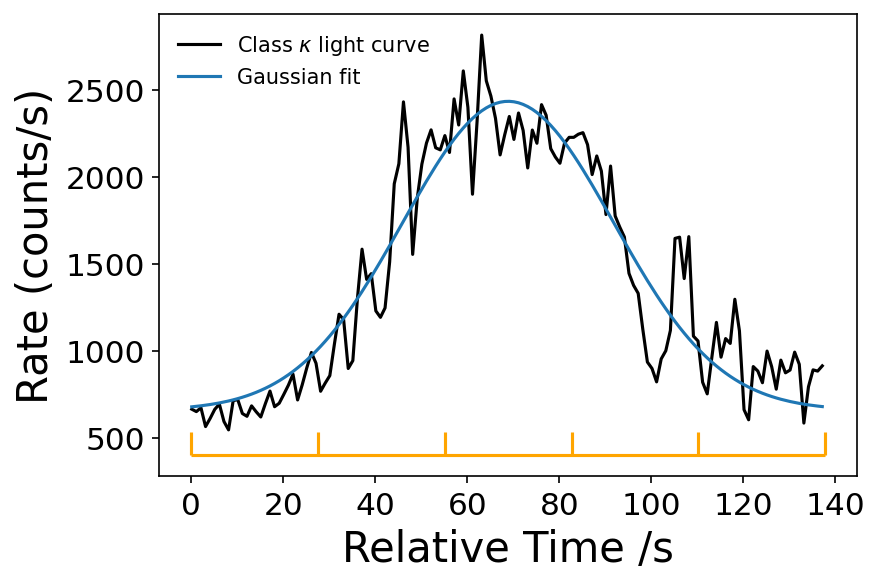}
    \caption{The Gaussian fit of class $\psi$ (left) and $\kappa$ (right) light curve to determine the interval of each main pulse and corresponding GTIs.}
    \label{fig:Gaussian}
    
\end{figure*}

To obtain the phase-resolved spectra for the class $\psi$, we find the local maximum value of the peaks in the light curves and fit them with a Gaussian function. For the class $\kappa$, we fit all the data points in each main pulse of light curves with a Gaussian function since there are few significant mini pulses, see Figure \ref{fig:Gaussian}. After obtaining the center of the main pulse $\mu$ and standard deviation $\sigma$ from the fitting results, we determine the interval of each main pulse ranging from $\mu$-3$\sigma$ to $\mu$+3$\sigma$ both for class $\psi$ and $\kappa$. Then we divide all these time intervals into 5 phase intervals: 0-0.2, 0.2-0.4, 0.4-0.6, 0.6-0.8, 0.8-1.0, showing as orange line segments in Figure \ref{fig:Gaussian} and generate the corresponding GTIs and folded spectra.

\begin{figure*}
    \centering
	\includegraphics[width=0.9\textwidth]{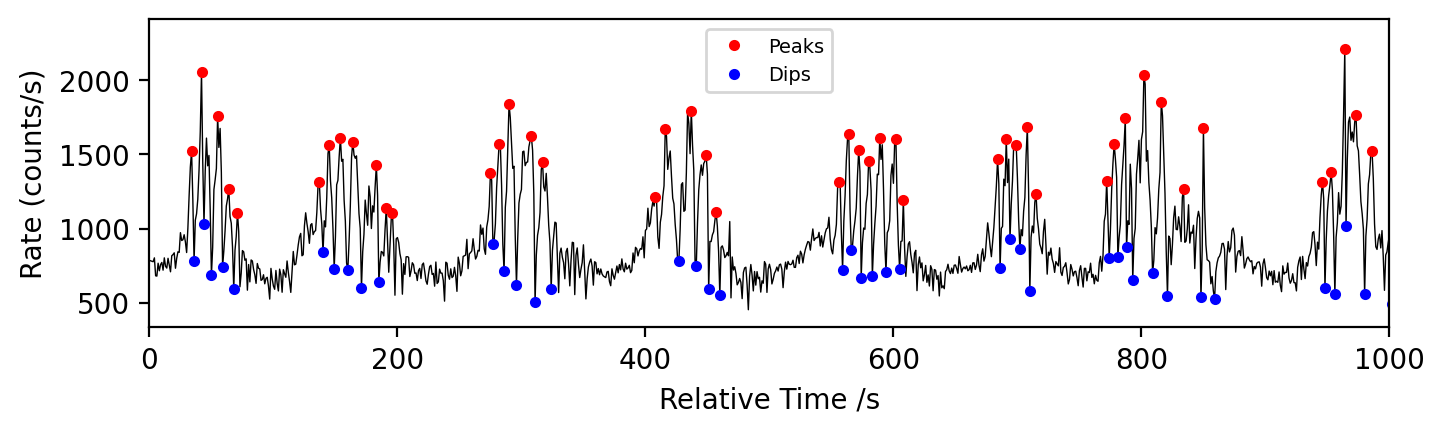}
    \caption{The peaks (red) and dips (blue) of mini pulses in class $\psi$ (obsID 0103010107). }
    \label{fig:peaks}
    
\end{figure*}

For the analysis of peaks and dips of mini pulses, we search the local maximum and minimum value of class $\psi$ light curves. In order to exclude the insignificant peaks or dips, we select the peaks or dips with prominence or depth of count rates over 500 and require the peaks to have count rates over 1000 in the light curves, as shown in \ref{fig:peaks}.

\section {PSD analysis}
\label{A2}
\begin{figure*}
	\includegraphics[width=\textwidth]{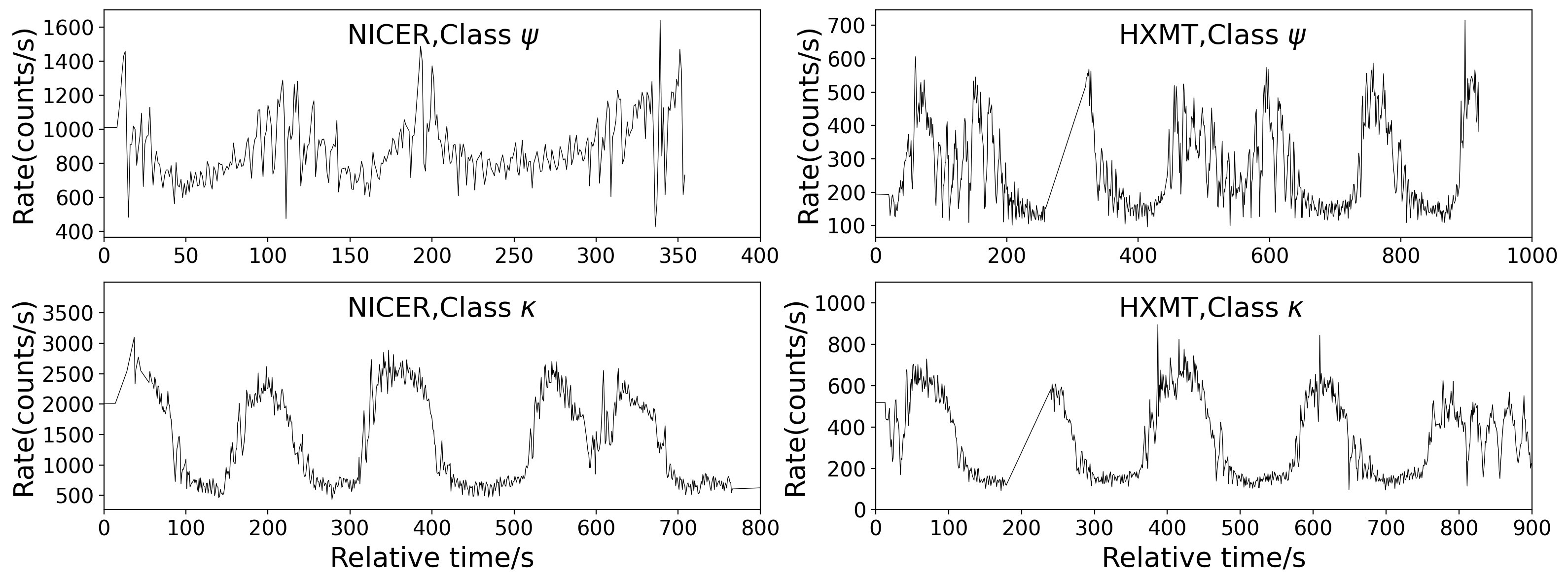}
	\caption{Examples of the light curves of class $\psi$ and $\kappa$ observations with short GTIs.}
	\label{fig:shortlc}
\end{figure*}

The observations of classes $\psi$ and $\kappa$ with GTI shorter than 1200 seconds, which are not included in our PSD analyses Table \ref{tab:psdtable} due to the low signal-to-noise ratio. We present two examples in Figure \ref{fig:shortlc}, where the IDs of these observations are shown in Table \ref{tab:nobID}.
\begin{table}
	\centering
	\caption{The observation IDs with GTI shorter than 1200 seconds.}
	\label{tab:nobID}
	\begin{tabular*}{0.26\textwidth}{cc}
		\hline
		$\psi$ & $\kappa$\\
		\hline
        0103010105&P010131000103\\
        0103010106&P010131000104\\
        P010133000104&P010131000105\\
        P010133000110&1103010105\\
        P010133000115&1103010107\\
         &1103010108\\
         &1103010113\\
         &1103010115\\
         &1103010116\\
		\hline
	\end{tabular*}
\end{table}

\begin{table*}
	\centering
	\caption{The observation ID for different class variabilities as observed by NICER and \textit{Insight}-HXMT. The obsIDs beginning with P are observations by \textit{Insight}-HXMT and the other are NICER observations.}
	\label{tab:obID}
	\begin{tabular}{ccccc}
		\hline
		\multicolumn{2}{c}{$\psi$}&$\kappa$&$\gamma$&$\nu$\\
		\hline
	0103010105&P010133000107&P010131000103&1103010119&1103010135\\
	0103010106&P010133000110&P010131000104&1103010120&1103010136\\
	0103010107&P010133000111&P010131000105& &P010133001501\\
	0103010108&P010133000112&1103010103& &P010133001502\\
	0103010102&P010133000113&1103010105& &P010133001503\\
	P010133000103&P010133000114&1103010106& &P010133001504\\
	P010133000104&P010133000115&1103010107& &P010133001505\\
	P010133000105&P010131000101&1103010108& &P010133001509\\
	P010133000106&P010131000102&1103010109& & \\
	 & &1103010112& & \\
	 & &1103010113& & \\
	 & &1103010115& & \\
	 & &1103010116& & \\
		\hline
	\end{tabular}
\end{table*}

\bsp	
\label{lastpage}
\end{document}